\documentclass{emulateapj}
\usepackage{amsfonts,amsmath,amssymb}
\usepackage{graphicx}
\usepackage{color}
\usepackage{url,hyperref}

\begin{document}

\title{Software Use in Astronomy: An Informal Survey}
\shorttitle{Astro Software Survey}

\author{Ivelina Momcheva\altaffilmark{1},
        Erik Tollerud\altaffilmark{1, 2}
       }
       
\altaffiltext{1}{Astronomy Department, Yale University, P.O. Box 208101, New Haven, CT 06510, USA; ivelina.momcheva@yale.edu, erik.tollerud@yale.edu}
\altaffiltext{2}{Hubble Fellow}


\begin{abstract}
We report on an informal survey about the use of software in the worldwide astronomical community. The survey was carried out between December 2014 and February 2015, collecting responses from 1142 astronomers, spanning all career levels. We find that all participants use software in their research. The vast majority of participants, 90\%, write at least some of their own software. Even though writing software is so wide-spread among the survey participants, only 8\% of them report that they have received substantial training in software development. Another 49\% of the participants have received ``little'' training. The remaining 43\% have received no training. We also find that astronomers' software stack is fairly narrow. The 10 most popular tools among astronomers are (from most to least popular): Python, shell scripting, IDL, C/C++, Fortran, IRAF, spreadsheets, HTML/CSS, SQL and Supermongo. Across all participants the most common programing language is Python ($67\pm 2\%$), followed by IDL ($44\pm 2\%$), C/C++ ($37\pm 2\%$) and Fortran ($28\pm 2\%$). IRAF is used frequently by $24\pm 1\%$ of participants. We show that all trends are largely independent of career stage, area of research and geographic location.
\end{abstract}


\section{Introduction}
\label{sec:intro}

\begin{figure*}[t!]
\begin{center}
\includegraphics[width=1.7\columnwidth]{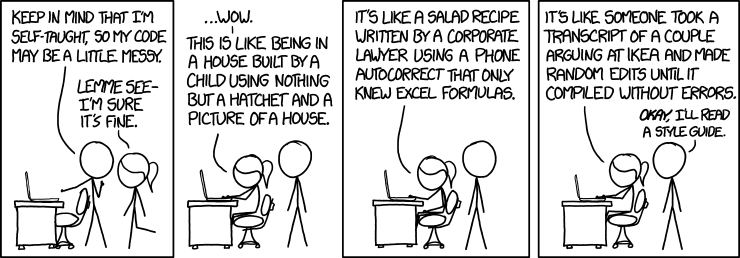}
\caption{ 
\label{fig:xkcd}
An \href{https://xkcd.com/1513/}{xkcd.com} comic strip that captures one of the problems with the lack of software training.  This strip was widely shared among astronomers on social networks, showing that the problem is known in at least part of the community.
}
\end{center}
\end{figure*}

Much of modern Astronomy research depends on software. Digital images and numerical simulations are central to the work of most astronomers today, and anyone who is actively involved in astronomy research has a variety of software techniques in their toolbox. Furthermore, the sheer volume of data has increased dramatically in recent years. The efficient and effective use of large data sets increasingly requires more than rudimentary software skills. Finally, as astronomy moves towards the open code model, propelled by pressure from funding agencies and journals as well as the community itself, readability and reusability of code will become increasingly important (Figure \ref{fig:xkcd}). Yet we know few details about the software practices of astronomers. In this work we aim to gain a greater understanding of the prevalence of software tools, the demographics of their users, and the level of software training in astronomy.

The astronomical community has, in the past, provided funding and support for software tools intended for the wider community. Examples of this include the Goddard IDL library (funded by the NASA ADP), IRAF (supported and developed by AURA at NOAO), STSDAS (supported and developed by STScI), and the Starlink suite (funded by PPARC). As the field develops, new tools are required and we need to focus our efforts on ones that will have the widest user base and the lowest barrier to utilization. For example, as our work here shows, the much larger astronomy user base of Python relative to the language R suggests that tools in the former language are likely to get many more users and contributers than the latter. 

More recently, there has been a growing discussion of the importance of data analysis and software development training in astronomy (e.g., the special sessions at the 225th AAS ``Astroinformatics and Astrostatistics in Astronomical Research Steps Towards Better Curricula'' and ``Licensing Astrophysics Codes'', which were standing room only). Although astronomy and astrophysics went digital long ago, the formal training of astronomy and physics students rarely involves software development or data-intensive analysis techniques. Such skills are increasingly critical in the era of ubiquitous ``Big Data'' (e.g., \citet{Berriman_2011}, or the \href{http://www.noao.edu/meetings/bigdata/}{2015 NOAO Big Data conference}). Better information on the needs of researchers as well as the current availability of training opportunities (or lack thereof) can be used to inform, motivate and focus future efforts towards improving this aspect of the astronomy curriculum. 

In 2014 the Software Sustainability Institute carried out an inquiry into the software use of researchers in the UK  (\cite{f824cd98-b953-4c08-96c8-2533188bc4c4}, see also \href{http://wl.figshare.com/articles/1243288/embed?show_title=1}{the associated presentation}). This survey provides useful context for software usage by researchers, as well as a useful definition of ``research software'':
\begin{quote}
Software that is used to generate, process or analyze results that you intend to appear in a publication (either in a journal, conference paper, monograph, book or thesis). Research software can be anything from a few lines of code written by yourself, to a professionally developed software package. Software that does not generate, process or analyze results - such as word processing software, or the use of a web search - does not count as ‘research software’ for the purposes of this survey.
\end{quote}
However, this survey was limited to researchers at UK institutions.  More importantly, it was not focused on astronomers, who may have quite different software practices from scientists in other fields.

Motivated by these issues and related discussions during the .Astronomy 6 conference, we created a survey to explore software use in astronomy.  In this paper, we discuss the methodology of the survey in \S \ref{sec:datamethods}, the results from the multiple-choice sections in \S \ref{sec:res} and the free-form comments in \S \ref{sec:comments}. In \S \ref{sec:ssicompare} we compare our results to the aforementioned SSI survey and in \S \ref{sec:conc} we conclude.

We have made the anonymized results of the survey and the code to generate the summary figures available at \url{https://github.com/eteq/software_survey_analysis}. This repository may be updated in the future if a significant number of new respondents fill out the survey\footnote{\url{http://tinyurl.com/pvyqw59}}.
    

\begin{figure}[]
\begin{center}
\includegraphics[width=1\columnwidth]{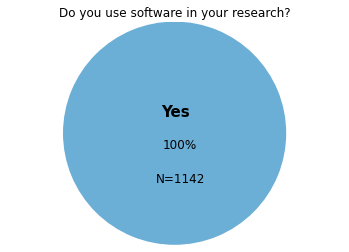}
\caption{ 
\label{fig:doyouse}
Responses to the question ``Do you use software in your research?''. 100\% of survey participants answered in the affirmative. 
}
\end{center}
\end{figure}

\section{Data and Methods}
\label{sec:datamethods}

The survey was constructed as a Google form questionnaire with seven questions and one comment box. Four of the questions were about software use and inspired by the SSI survey:
\begin{enumerate}
\item Do you use software in your research?
\item Have you had formal training in software development?
\item Which of these is more common in your work? (on writing one's own software)
\item Select any of these that you use regularly to write code for your research. (on most commonly used software tools)
\end{enumerate}
The remaining three questions requested basic demographic information:
\begin{enumerate}
\item What is your field of research?
\item What is your career stage?
\item What is the location of your institution?
\end{enumerate}

The survey was opened on December 9, 2014. The attendees of the .Astronomy 6 conference were asked to forward a link to the survey to their home departments, including a prompt to send it on to any other interested astronomer groups. A link to the survey was also posted on the Astronomers Facebook group. The survey received 758 responses on the first day and another 210 during the following day. The data for this work was collected on February 3, 2015. The number of participants at that time was 1145. Three responses from participants who indicated that they work in fields other than astronomy were removed for a final tally of 1142 participants.

\subsection{Survey Demographics}

The demographics of the sample at the time of collection were the following. Of the 1142 participants, 380 are graduate students, 340 are postdocs, 385 are research scientists and faculty (175 and 200, respectively). The remaining 37 were undergraduate students (10), emeritus professors, support scientists at observatories, adjunct faculty, post-bachelor's researchers, etc. These 37 ``miscellaneous'' career levels will be included in the analysis of the full sample but will not be included in any of the other career subgroups. For the analysis, we combine the research scientist and faculty subgroups to create three groups of similar size, roughly corresponding to ``early'', ``intermediate'' and ``late'' career stages.

In terms of areas of research, 823 participants chose ``Observational Astronomy/Astrophysics'', 353 selected ``Theoretical Astronomy/Astrophysics'', 130 indicated that they work in astronomical instrumentation and 66 in planetary science. 22 participants did not choose any of these four main categories. Of these, nine participants selected ``Other'', three did not choose an area of research, and 10 entered custom values such as physics, astro-statistics, cosmology, astroparticle physics, space physics, etc. Participants were allowed to choose more than one area of research, which is why the numbers for the individual categories add to more than 1142. 

The final piece of demographic data we collected was the geographic location of the participants' home institution. The majority of participants are from the USA (546), followed by Germany (170), UK (90), Australia (69) and Chile (35). 80\% of the participants come from these five top countries, with 48\% from the USA. The remaining 232 participants come from 41 different countries. The break down is the following:

             Netherlands     (31),
                  Sweden     (21),
               Argentina     (19),
                  Canada     (18),
                  Brazil     (14),
                  France     (13),
                   Spain     (12),
                   Italy     (11),
                  Poland      (9),
                  Mexico      (9),
             Switzerland      (9),
                  Israel      (6),
                 Denmark      (5),
                 Finland      (3),
                   India      (3),
                 Ireland      (3),
                Portugal      (3),
                   Japan      (3),
    United Arab Emirates      (2),
             South Korea      (2),
            South Africa      (2),
                  Russia      (2),
             South Korea      (2),
                 Belgium      (2),
                 Austria      (2),
             New Zealand      (2),
                  Greece      (1),
               Lithuania      (1),
                 Georgia      (1),
                Malaysia      (1),
                  Norway      (1),
                Slovakia      (1),
          Czech Republic      (1),
                   China      (1),
               Swaziland      (1),
                  Taiwan      (1),
                  Turkey      (1),
              Uzbekistan      (1),
                 Hungary      (1),
                Vatican       (1),
                   Ghana      (1).
                   
The geographic distribution of the participants indicates that our methods of circulating the survey were unable to reach a broad base of researchers in Asia, Africa, and Eastern Europe. Compared to the \href{http://www.iau.org/administration/membership/individual/distribution/}{IAU membership}, the USA is over-represented by a factor of 2.1, Germany by a factor of 2.7, and China is under-represented by a factor of 60. Major contributors to this imbalance are the language of the survey (English) and the method of distribution (social media and friends-of-friends networks). Hence, any conclusions we make will be only applicable to the researchers working in the countries which are predominantly represented.

\subsection{Survey Completeness}

This survey should \emph{not} be viewed as a systematically representative sample of the astronomy community.  
The request to fill out the survey was primarily spread via social networks (i.e., departmental e-mails, Facebook, twitter, etc.), so it is possible that the sample of astronomers surveyed is biased in ways that may affect the results presented below.
Given that we are astronomers (not social scientists), we are not trained in methods to address these effects, and therefore simply present the raw results of the survey with no correction for selection bias.
That said, the sheer number of responses implies these results represent a significant part of the community.
Regardless, we would certainly be happy if this work inspires a more rigorous survey by social scientists with domain-specific expertise.


\begin{figure*}[]
\begin{center}
\includegraphics[width=1.8\columnwidth]{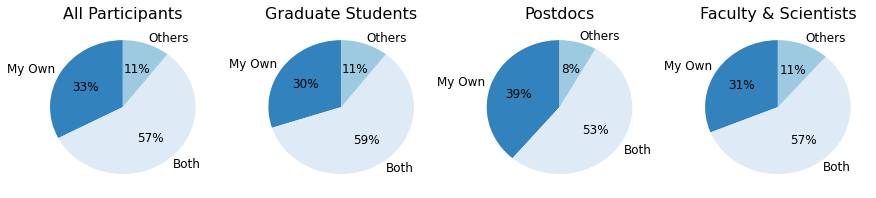}
\caption{ 
\label{fig:write1}
Answers to ``Which of these is more common in your work: I write mostly my own software, I mostly use software written by others, or somewhere in between'', sub-divided by career stage.
}
\end{center}
\end{figure*}

\begin{figure*}[]
\begin{center}
\includegraphics[width=1.8\columnwidth]{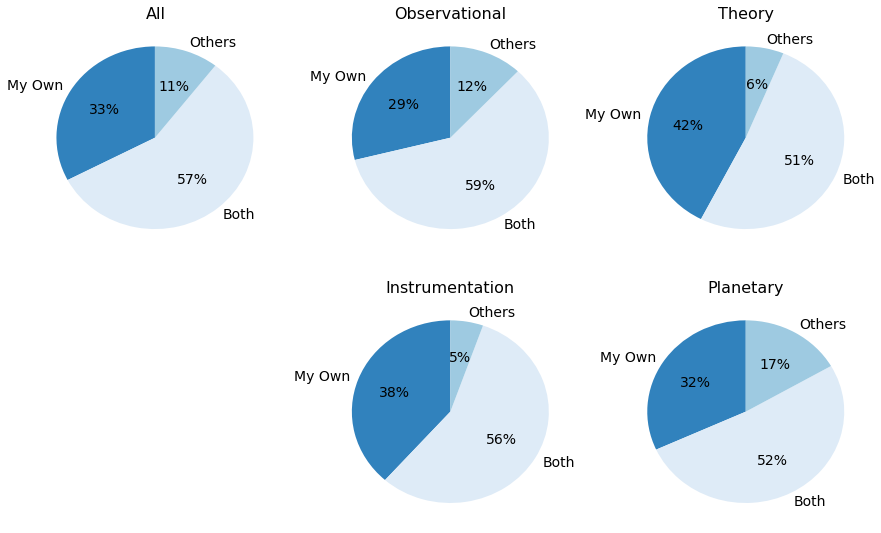}
\caption{ 
\label{fig:write2}
Same question as Figure \ref{fig:write1}, but divided by sub-field.
}
\end{center}
\end{figure*}

\begin{figure*}[]
\begin{center}
\includegraphics[width=1.8\columnwidth]{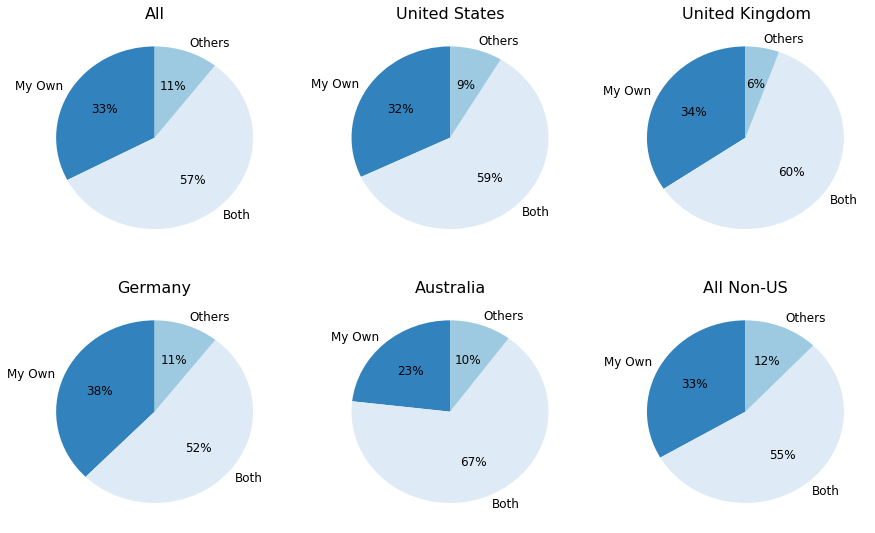}
\caption{ 
\label{fig:write3}
Same question as Figure \ref{fig:write1}, but sub-divided by country (for the countries with the highest number of respondents).
}
\end{center}
\end{figure*}

\section{Results}
\label{sec:res}

In this section we show and describe the results from the survey. We first focus on software use and whether astronomers write their own software. We then examine the training we receive in software development. Finally, we discuss the most commonly used software tools. In all cases we consider how career stage, research area and geographic location alter these results. Where relevant, we assume Poisson statistics for error bars and to provide estimates of significance.

\subsection{Software Use}

The first question of the survey aims to establish a baseline of software use. The answers to the question ``Do you use software in your research?'' are shown in Figure \ref{fig:doyouse}. Unanimously, all participants responded with ``Yes''. This unanimity is not surprising: It would be difficult to imagine pursuing astronomical research today which does not rely on software at least to some extent. But the unanimity does serve to underscore the importance of software in the field.


\begin{figure*}[]
\begin{center}
\includegraphics[width=1.8\columnwidth]{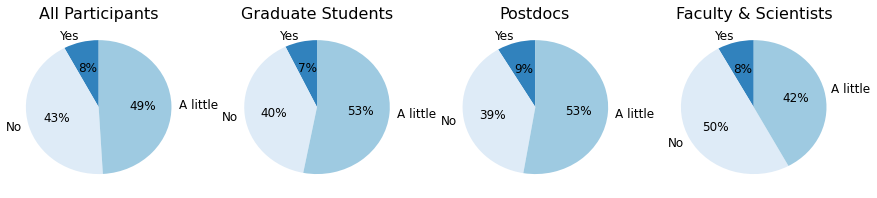}
\caption{ 
\label{fig:train1}
Answers to the question ``Have you had formal training in software development: Yes, a lot; Yes, a little; No'', sub-divided by career stage.
}
\end{center}
\end{figure*}

\subsection{Do Astronomers Use Their Own Software?}
\label{ssec:own}

We ask the survey participants to best describe the authorship of the software that they use. The goal of this question is to find what is the predominant practice in the community: do most of us use ``black box'' software packages written by few or do most of us write custom software? We find that most often we do both (Figure \ref{fig:write1}, first panel): $57\pm2\%$ choose this option. One third of survey participants say they mostly write their own software ($33\pm2\%$), while only a small portion of survey participants predominantly use software written by others: $11\pm1\%$. Overall, 89\% of all participants write some of their own software.

In Figure \ref{fig:write1} we also explore the breakdown of the answers as a function of career stage. The answers vary slightly between the three groups. One curious result is that the answers of the ``early'' and ``late'' career stages closely resemble each other. The reason for such a trend may be that students follow their advisors' recommendations on software practices. In both of these groups, $\sim30\pm3\%$ predominantly write their own software, while $\sim11\pm2\%$) mostly use software written by others. In contrast, a larger portion of postdocs write their own software: $39\pm3\%$. 

We further consider the breakdown of answers as a function of research area in Figure \ref{fig:write2}. The groups are not fully independent because participants were allowed to choose more than one research area. In all groups the largest portion of astronomers, $\ge50\%$, use both software written by others and write their own. Researchers working in theory and instrumentation are more likely to primarily depend on their own software ($42\pm4\%$ and $38\pm5\%$, respectively) than planetary and observational astronomers ($32\pm7\%$ and $29\pm2\%$, respectively). The latter two groups are more likely to primarily use software written by others, $17\pm5\%$ and $12\pm1\%$ for planetary and observational researchers, respectively, versus $6\pm1\%$ and $5\pm2\%$ for theory and instrumentation.

Finally, we break down the answers by country of the researchers' home institution as shown in Figure \ref{fig:write3}. These and the following plots by country are more difficult to interpret because they only contain information about the researchers' current institutions rather than their institutional history. All countries show similar trends, with $\ge50\%$ of astronomers choosing the ``Both'' option. At the extremes, the survey respondents from Germany are most likely to write their own software, $38\pm5\%$, and the respondents from the UK are least likely to use software written by others, $6\pm3\%$.

In conclusion, the majority of astronomers, $\sim90\%$ write at least some of their own software, across all demographics explored in our survey and a third of survey participants predominantly rely on their own software.  


\begin{figure*}[]
\begin{center}
\includegraphics[width=1.8\columnwidth]{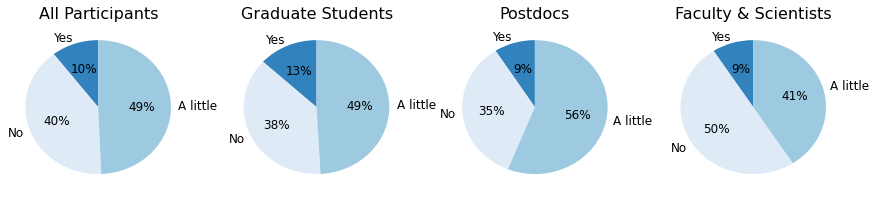}
\caption{ 
\label{fig:crosstrain}
Answers to the question from Figure \ref{fig:train1}, but only for those who primarily write their own software.
}
\end{center}
\end{figure*}

\begin{figure*}[]
\begin{center}
\includegraphics[width=1.8\columnwidth]{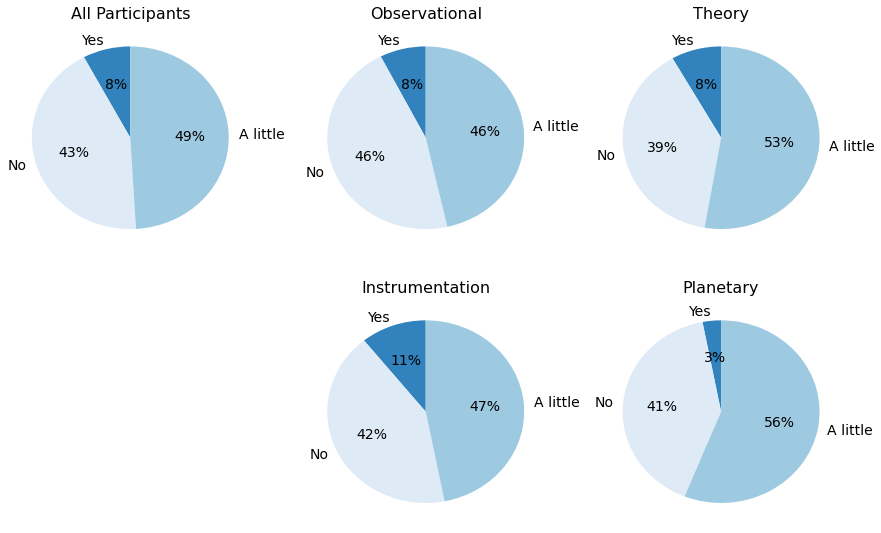}
\caption{ 
\label{fig:train2}
Same question as Figure \ref{fig:train1}, but sub-divided by sub-field.
}
\end{center}
\end{figure*}

\begin{figure*}[]
\begin{center}
\includegraphics[width=1.8\columnwidth]{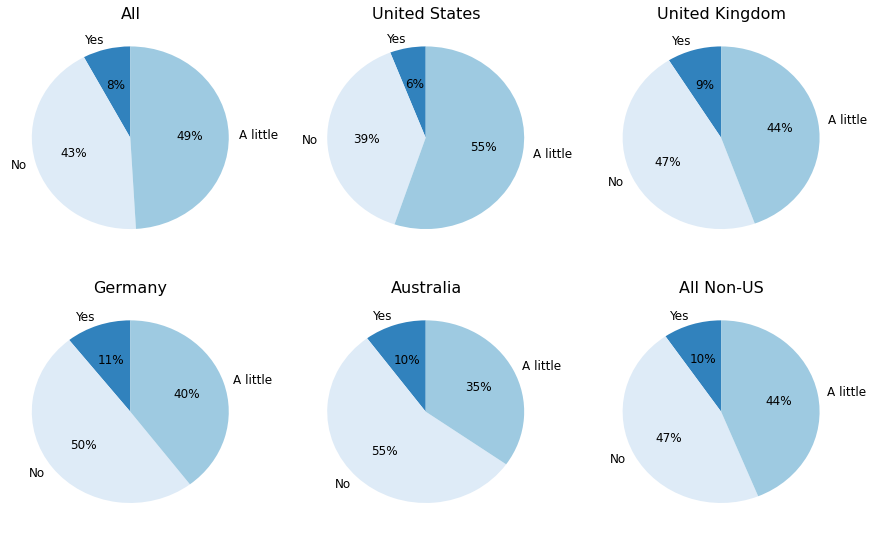}
\caption{ 
\label{fig:train3}
Same question as Figure \ref{fig:train1}, but sub-divided by country.
}
\end{center}
\end{figure*}

\subsection{Are We Trained?}

Considering that, across all demographics, $\sim90\%$ of astronomers are involved in writing software (\S \ref{ssec:own}), it is important to asses the level of training we receive. We allowed participants to choose from one of three levels of training in software development: ``A little'', ``A lot'', or ``None''. The survey questions did not give guidelines on how to interpret the first two categories. Rather, we left it to the survey participants to decide whether they thought their training was substantial or not. This section breaks down the answers into different demographics. 

The first panel in Figure \ref{fig:train1} shows the answers from all participants. Overall, $8\pm1\%$ of survey participants have received substantial training, $49\pm2\%$ have received a little training and $43\pm2\%$ have received no training. Altogether, $57\pm2\%$ of survey participants have received some training in software development. Across all career levels, only $\sim8\%$ of astronomers have received significant training. Facutly and scientists are slightly more likely to have received no training at $50\pm4\%$ versus $40\pm3\%$ for the more junior groups. Postdocs and graduate students are slightly more likely to have received some training at $53\pm4\%$, relative to faculty and scientists ($42\pm3\%$).

In Figure \ref{fig:crosstrain} we specifically focus on the training of survey participants who, in the previous questions, said that they primarily write their own software (``My own'' option, $\sim33\%$ of the sample).  Overall, $40\pm3\%$ of those participants have received no training and $89\pm5\%$ have received at best a little bit of training. The results for this subset are consistent with the answers from the full sample within the error bars, i.e. astronomers who primarily write their own software do not have more training in software development than everyone else. The results are similar if we also considered the participants who write some of their software (``Both'' option). This finding is key because it shows that the lack of training is not because there is no need for such skills. Rather training simply does not occur. This is of particular importance because it implies that many astronomers have little to no training in an activity that is a major part of their research work, despite the fact that they nearly always have many years of post-secondary education during which they could have received such training.

Back to the full sample, in Figure \ref{fig:train2} we show the breakdown of answers as a function of research area. The trends remain the same across all fields.
The breakdown by country (Figure \ref{fig:train3}) shows that the results are similar internationally. The fraction of astronomers with significant training is largely independent of geography. Some geographical variations exist in the fraction of participants who have at least a little training: the USA has the largest fraction with training: $55\pm3\%$, while Australia has the smallest with $35\pm7\%$. Based on these results, we speculate that opportunities to receive at least a little bit of training are more available at US institutions or that more US researches seek out such opportunities. 

In conclusion, across all career levels, research areas and countries, only a small fraction of astronomy researchers receive significant training in software development. The lack of a strong trend with career level may indicate that significant training only occurs at the undergraduate level (and some participants left comments to that effect). While graduate students are more likely to have had a little training, it seems that few graduate programs offer and/or require CS courses (otherwise junior astronomers will have a higher level of significant training). Overall, $\sim90\%$ of the survey participants have received only a little bit of training at best, despite \emph{all} being software users, and most being writers of their own software.


\begin{figure*}[]
\begin{center}
\includegraphics[width=2.0\columnwidth]{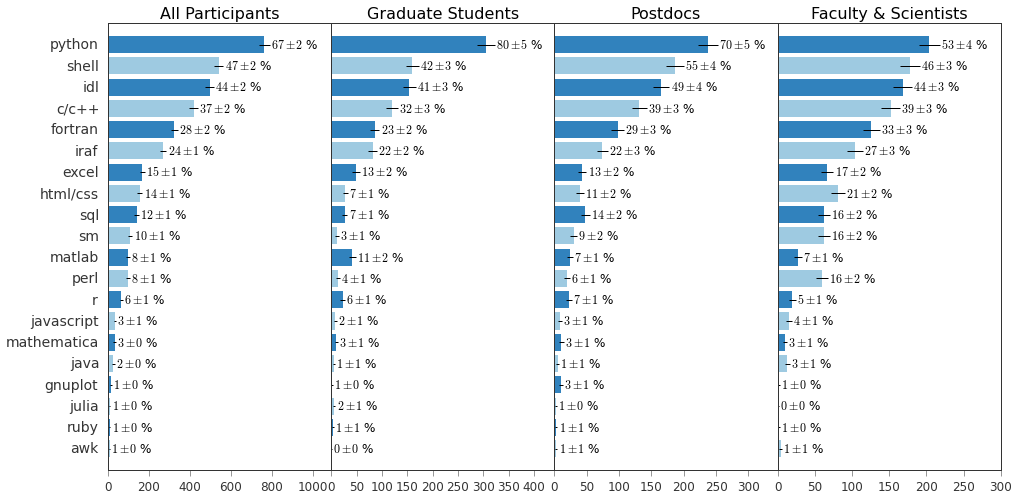}
\caption{ 
\label{fig:stack1}
Responses to the prompt ``Select any of these that you regularly use in your research'', sub-divided by career stage.  The options listed included: IDL, IRAF, Python, C, Fortran, Perl, Javascript, Julia, Matlab, Java, R, SQL, Shell Scripting, STAN, Figaro, Ruby, HML/CSS, Supermongo (labeled ``sm''), and Excel or other spreadsheets (labeled ``excel'').  Respondents could add additional tools not listed using an ``Other'' box. Among the tools in this plot, four items were added by respondents: C++, Mathematica, gnuplot and awk. Note that the x axis varies between panels.
}
\end{center}
\end{figure*}


\subsection{What is in the Astronomer Software Tool Stack?}

In this section we consider the most common software tools for professional astronomers. We refer to the full set of software tools an astronomer uses as their ``stack''. In the survey form we suggested 19 software tools and allowed participants to add any options we missed. The input was edited to standardize spelling and capitalization of tools. In total, participants added 64 custom options. 10 respondents did not provide an answer to this question. While ``C'' was an option, ``C++'' was not part of our suggestions. Some participants noted in the comments what they chose ``C'' even though they actually use ``C++''. For this reason we consider C and C++ together in our analysis. Within the top-20 most used software tools there are four items that were not on our original list: C++, Mathematica, gnuplot and awk.

The overall astronomer stack is rather narrow (Figure \ref{fig:stack1}, first panel). Only ten of the software tools are used by more than 10\% of the survey participants. These are (from most popular to least popular): Python, shell scripting, IDL, C/C++, Fortran, IRAF, spreadsheets, HTML/CSS, SQL and Supermongo. Across all participants the most common programing language is Python ($67\pm2\%$), followed by IDL ($44\pm2\%$), C/C++ ($37\pm2\%$) and Fortran ($28\pm2\%$). Shell scripting is the second most popular tool for astronomers ($47\pm2\%$).  The IRAF (Image Reduction and Analysis Facility) environment is used by $24\pm1\%$ of the survey participants. 

Across the different career stages, we notice that senior astronomers have a broader tool stack, i.e. they utilize a wider variety of tools in their research. Only eight tools are used by more than 10\% of graduate students, nine tools are used by more than 10\% of postdocs and 11 tools are used by more than 10\% of faculty and scientists. Python is the most popular tool at all career levels, and it is most popular among junior researchers. Four out of five graduate students use Python ($80\pm5\%$), as do $70\pm5\%$ of postdocs and half of faculty and scientists ($53\pm4\%$). IDL, IRAF and compiled languages have a more uniform user base across all career levels. Some tools are unique to certain demographics. Graduate students have the highest fraction of Matlab users ($11\%$), while faculty and research scientists dominate HTML/CSS (21\%), Supermongo (16\%) and Perl (16\%).

\begin{figure*}[]
\begin{center}
\includegraphics[width=2.0\columnwidth]{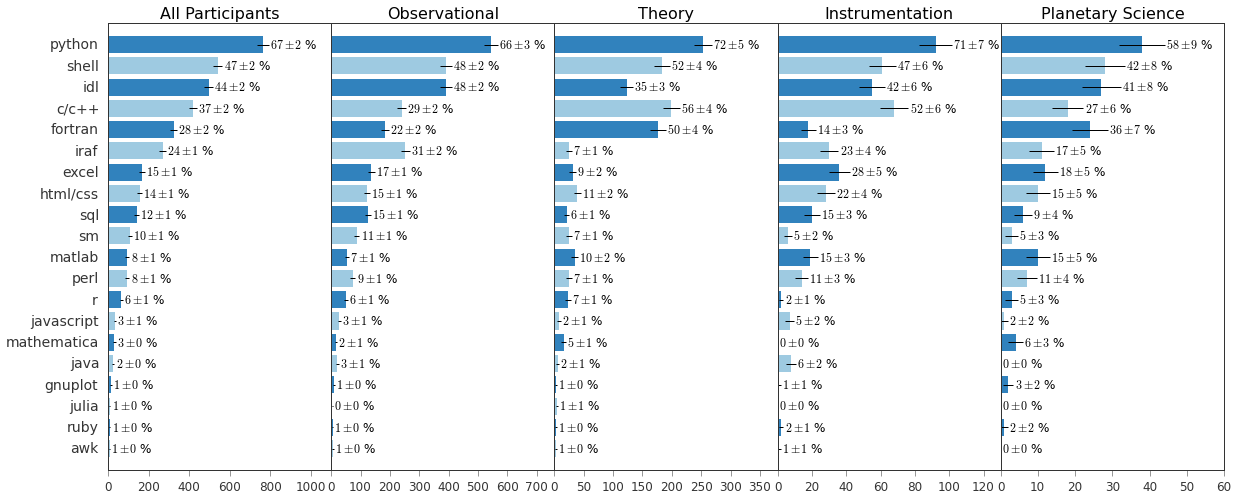}
\caption{ 
\label{fig:stack2}
Same question as Figure \ref{fig:stack1}, but sub-divided by sub-field.
}
\end{center}
\end{figure*}

Unsurprisingly, software tools depend strongly on the research area (Figure \ref{fig:stack2}). Without attempting to be exhaustive, we note some interesting differences between fields. Observational astronomers have the highest fractions of IDL ($48\pm2\%$) and IRAF ($31\pm2\%$) users. Theoretical researchers have the highest fractions of compiled language users: C/C++ with $56\pm4\%$ and Fortran with $50\pm4\%$. Researchers in instrumentation have a high fraction of C/C++ ($52\pm6\%$) and spreadsheet ($28\pm5\%$) users. Other tools, however show little field-to-field variation. Python use is consistently high across all fields at 60 - 70\%, as is shell scripting at $\sim50\%$. 

\begin{figure*}[]
\begin{center}
\includegraphics[width=2.0\columnwidth]{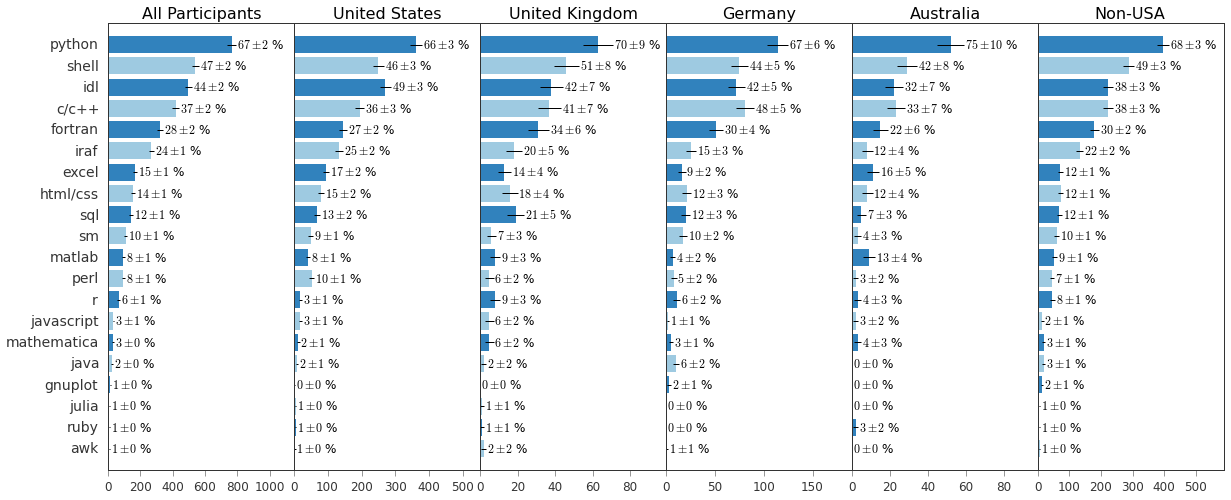}
\caption{ 
\label{fig:stack3}
Same question as Figure \ref{fig:stack1}, but sub-divided by country (for the countries with the highest number of respondents).
}
\end{center}
\end{figure*}

Finally, in Figure \ref{fig:stack3} we consider the software stack for researchers in different countries. Researches in the USA have the highest fractions of IDL ($49\pm3\%$) and IRAF ($25\pm2\%$) users, while Australia has the lowest fraction of users of these tools, $32\pm7\%$ and $12\pm4\%$, for IDL and IRAF respectively. The UK has the highest fraction of SQL users ($21\pm5\%$); Germany has the highest fraction of C/C++ users ($48\pm5\%$); and Australia has the highest fraction of Matlab users ($13\pm4\%$). However, these results can be strongly influenced by the research areas represented for each country within our sample so we caution against drawing far-reaching conclusions. 

We can also compare the USA and non-USA survey respondents, since those two samples are comparable in size (Figure \ref{fig:stack3}, second and sixth panels). Overall the rankings and fractions of users of different tools are very similar as can be expected by the global mobility of many astronomers. The only notable exceptions are IDL and R. The fraction of IDL users in the USA is 10\% larger than of non-USA participants. The user base of the statistical package R is reversed:  $8\pm1\%$ of non-USA researchers choose this option vs. only $3\pm1$ of USA researchers. Considering the wide-spread use of R in other scientific fields, its popularity among astronomers is strikingly low.


\subsection{Python vs. IDL?}

\begin{figure}[]
\begin{center}
\includegraphics[width=1\columnwidth]{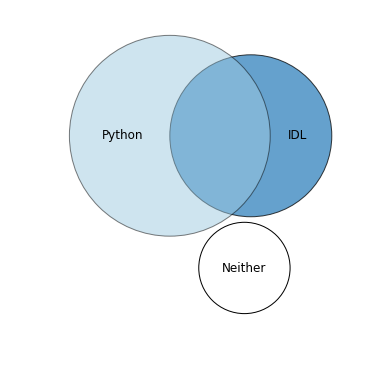}
\caption{ 
\label{fig:venn}
Venn Diagram of the overlap between respondents who reported using Python, IDL, or neither.
}
\end{center}
\end{figure}

A recent shift in astronomy has been the favored choice of interpreted programming language for day-to-day analysis work. In the previous section we showed that Python has overtaken IDL in popularity. This may not have been true three to five years ago, but today Python is, by a wide margin, the most popular interpreted language in astronomy (at least insofar as this survey is representative). Still, there is a significant overlap between the users of both languages as many people are either transitioning from one to the other or using both in their research. In Figure \ref{fig:venn} we show a Venn diagram of the Python and IDL users. In total 984 (86\%) of the survey participants use either Python or IDL. Of those, 764 use Python and 497 use IDL. Both are chosen by 277 or 25\% of all survey participants. This indicates substantial overlap: 36\% of Python users also use IDL and 55\% of IDL users also use Python. Finally, 158 survey participants (14\% of the full sample) chose neither option.


\subsection{Interactive Visualization Of Software Tools}
\label{ssec:d3viz}

To facilitate understanding of this multi-dimensional dataset of how use of the various software tools overlap with each other, we provide an interactive visualization, available within the \href{https://www.authorea.com/users/10533/articles/18046/_show_article}{Authorea version of the paper}, by downloading the software repository described in Section \ref{sec:intro}, or at \href{http://eteq.github.io/software_survey_analysis/software_tools_heirarchy_d3vis.html}{this link}. In this visualization, the tools respondents use are shown as sectors in a radial layout.  Users of multiple tools are represented as stacked sectors: for example, the fraction of users who use only Python and IDL are represented as the fraction of the third ring labeled ``idl'' with ``python'' and ``idl'' as the lower two layers.  Hovering over that sector shows the number of respondents to the left of the page (for Python and IDL only it is 36.)  Clicking on a particular sector expands the visualization to show only those who use the corresponding stack of tools, while clicking on the central circle goes back to larger segments of the survey.


\section{Comments}
\label{sec:comments}

We allowed participants to leave comments at the end of the survey. In order to increase the anonymity of comments we detached them from the answers to the other questions (aside from career stage). We further removed e-mails, names, or other identifying information from the content of the comments. If anyone who took the survey would prefer that their comments not be included, we ask them to contact us and we would be happy to remove the information from the public dataset. 

We see three recurring topics in the comments. The first common comment topic is the switch from IDL to Python. Many users comment that they would like to or are planning to make the switch from IDL to Python, frequently because of licensing issues and costs. We find it particularly striking that several senior astronomers commented that they are learning Python to be more helpful to their students.

The second common comment topic is the desire for more opportunities to improve software development skills. Many participants voiced interest in attending software development classes for astronomers. Several suggested that classes in programming and statistics should be an integral part of the undergraduate and/or graduate curriculum in astronomy. Participants suggested that such training not only aids in better research efficiency, but also can make code more readable and reusable. 

The third recurring topic in the comments is the lack of career opportunities for astronomers who write software. Comments suggest that more professional recognition should be given to those who spend most of their time developing important tools for the astronomical community and that such efforts should be recognized in hiring and explicitly funded. 

We cite some of the representative comments on each topic in the Appendix.


\section{Comparison to SSI Survey}
\label{sec:ssicompare}

This work was inspired by a UK survey led by the Software Sustainability Institute. While their findings are for the wider scientific community in the UK and ours are for the worldwide astronomical community, a comparison is still interesting. 

The SSI survey finds that 90\% of UK scientists use software in their research. Our survey shows that astronomy is in line with other sciences in the use of software --  100\% of astronomers use it. 90\% of astronomers write some software (93\% of UK astronomers). This is much larger than the SSI survey, which finds that only 56\% of researchers, across all disciplines, write some of their software.  This implies astronomers are much more dependent on their own software than other sciences.

In the SSI survey, 55\% of respondents say they have received some training in software development, with 40\% indicating that the training was a formal course and 15\% indicating self-directed study. Our categories are not identical, but we find that a similar fraction of astronomers -- 57\% (53\% of UK astronomers) -- say that they have received some form of training. However, only 8\% say that the training was substantial, while 49\% say they received a little training. The decision what constitutes a lot and a little was left to the participants. Those who chose to expand on their decision indicated that a lot corresponded to a formal class while a little corresponded to using on-line materials such as Software Carpentry and Code Academy.

Finally, 40\% of our survey respondents who predominantly write their own software have received no training (42\% of those who write some of their own software). This fraction is twice as large as the one reported in the SSI survey (21\%) and may indicate that in astronomy there are fewer efforts and opportunities to train researchers in software development.  This is all the more surprising given that many {\it more} astronomers write their own code, according to these surveys.



\section{Conclusions}
\label{sec:conc}


Based on the responses summarized in Section \ref{sec:res}, we come to the following conclusions:

\begin{itemize}
\item Unsurprisingly, all 1142 survey participants say that they do use software in their research (Figure 2). The unanimous answer to this question underscores the importance of understanding how astronomers use software (i.e., the purpose of this survey).
\item $89\%$ of astronomers across all demographics write their own software (Figure \ref{fig:write1}).  However, only 58\% of those who write software are trained in software.  Moreover, only 8\% self-report as having better than ``a little'' training.  From this we conclude that 42\% of astronomers have no training for a key element of their work, and 92 \% have at most ``a little'' training.  
\item Python is the dominant language among our respondents.  Surprisingly, this is true across all career stages (Figure \ref{fig:stack1}).  While a commonly-expressed mindset, reflected in some of the comments, is that graduate students are more likely to know the newer languages, it appears that this is only mildly true, at least in our survey. 
\item Astronomers have a fairly narrow software ``stack'', with only 10 tools used by more than 10\% or respondents. Theorists tend to have a  more narrow stack relative to other fields (Figure \ref{fig:stack2}) as do graduate students relative to more senior researchers. Independent of career level and field, Python and shell scripting are are most popular tools for astronomers. These results show that training efforts can have a significant impact even if they only focus on a limited number of software tools. We suggest that the rankings we produce can help in choosing training topics that would be most useful for the broadest group of participants. 
\end{itemize}

We caution that these results are tentative because our sampling methodology was not robust.  If nothing else, we hope this survey will prompt a more formal study of software use in astronomy to better understand how we should use the limited resources of our community to improve software training and software use.



\section{Acknowledgements}

We thank the organizers and participants of the .Astronomy 6 conference, the Adler Planetarium, and all participants in the survey.  Without them, this work would not have been possible. We also thank Demitri Muna and Camille Avestruz for helpful comments on the manuscript.


\begin{appendix}
\section{Participant Comments}

The section below shows a representative sampling of free-form comments left by the survey participants. 

On switching from IDL to Python:

\begin{itemize}
\item{I recently switched from using IDL as my primary programming language to using Python. (Graduate student)}
\item{Mainly IDL user who wants to switch to Python as it is more open source. (Graduate student)}
\item{I learned IDL as an undergrad and continue to mostly code in it as a grad student. However I've been learning Python lately and plan to mostly switch over within the next year or so. (Graduate student)}
\item{I plan to learn Python but haven't yet worked with it. (Graduate student)}
\item{At this stage I see Python as the future and am rapidly moving away from IDL. (Graduate student)}
\item{I learned IDL as an undergrad (class of 2004) and used it nearly exclusively [...] until about two years ago. Over the last two years I've been slowly switching to Python [...]
(Postdoc)}
\item{I've only recently started working in IDL and Python.  I expect to do quite a bit of development in Python from now on. (Postdoc)}
\item{I want to learn Python and R as soon as possible (Postdoc)}
\item{While I haven't learned it yet many of my colleagues use Python and I make all of my students learn that (instead of e.g. Matlab). (Faculty)}
\item{I am telling all of my students to learn Python and through that I am also gaining proficiency in Python.  This is different from what my advisor did.  He told me to use the language that he used so that he could help me debug it. (Faculty)}
\item{Moving to Python as IDL needs a license... And I just like the language. (Faculty)}
\item{Currently I code in IDL but I am trying to switch to Python. (Faculty)}
\item{I plan to switch from IDL to Python over the next 2-3 years. (Faculty)}
\item{I'm intending to try out Python soon. (Faculty)}
\item{Taking a workshop on Python soon I can encourage and help my students learn a language that can be used outside of academia as well. (Faculty)}
\end{itemize}

On the desire for more opportunities to improve software development skills:

\begin{itemize}
\item{If there is any software development training program for astronomers I would love to attend. (Graduate student)}
\item{I think it should be strongly recommended that people going into astronomy should take programming classes. In fact I would make it part of the required course work to get a B.S. in Physics or Astronomy. Most astronomers I know did not take any formal programming course we learned as we went along. Most of us write our own code and many do not use good coding practices making reading or adapting code from other astronomers a lot more painful than it should be. (Graduate student)}
\item{Helps to include courses in computation and statistics in grad curriculum. (Graduate student)}
\item{Astro students should get more formal training in programming and software design (and it should happen at the undergraduate level whenever possible). (Graduate student)}
\item{I wish I could get more formal training on programming. I have the feeling that Astronomy and Physics department usually don't emphasize the importance of programming until people start doing researches. (Graduate student)}
\item{The coding skills incoming graduate students possess seem to vary wildly but they are often dismal with respect to the level required to begin doing serious research right away. In my department there seems to be little motivation to rectify this with either: (1) requiring undergraduate CS preparation as a condition for admission; or (2) organizing a programming course for beginning graduate students. This situation seems if not unsustainable very much non-optimal for the cultivation of strong substantive independent research skills. I imagine many other departments are currently facing the same dilemma. (Graduate student)}
\item{More formal training in software use/development would be wonderful. (Graduate student)}
\item{Formal training on astronomical software packages as part of my astronomy undergraduate degree would have been very helpful for me. (Graduate student)}
\item{We need to be teaching undergraduates and graduates good object-oriented design skills from day one. Software is more important now than it ever was and the ``learn as you go'' mentality causes a tremendous amount of wasted efforts as bad code has to be rewritten all too often with the side effect of having astronomers with career skills that aren't as well developed as they could have been should they choose to leave astronomy. Also we should make a concerted effort to rewrite some of our legacy tools (IDL libraries, IRAF, etc.) in a language and style that is more easily and cleanly extended and maintained. Incentives to put in the time and money for these [initially] low impact projects are hard to come by though. (Graduate student)}

\end{itemize}

On the lack of career opportunities for people who write software:

\begin{itemize}
\item{People who develop software that the community use should be recognized more for their efforts! (Research Scientist)}
\item{It would be great to have more career options for researchers who focus on software development in the astrophysical community.  Too many good researchers who contribute lots of great software to the community have been forced out of the field because of lack of recognition for their work and lack of funding for people other than those who publish several science papers a year. (Postdoc)}
\item{Software development is evidently as important a tool in modern science as mathematics and just as it has historically not been deemed wise to outsource all mathematics to professional mathematicians I believe a large fraction of scientific software development will have to be accomplished by scientists who are intimately familiar with the problem at hand. Perhaps more than is the case for mathematics though the paper metric used for hiring scientists often pushes excellent software developers out of science and into industry who are then lost to us. (Postdoc)}
\item{I find that my observational colleagues are often unaware that we as computational scientists need to write proposals for supercomputers like they write for telescopes.  Also when we write science proposals (e.g., NASA, NSF) we have to lie about how much time we will spend developing code say only a few months when in reality it occupies most of the grant period since code development is frowned upon (except within the new NASA ROSES program PDART started in 2014). (Research Scientist)}
\end{itemize}

\end{appendix}







\end{document}